\documentclass[twocolumn,showpacs,floatfix,aps]{revtex4}
\usepackage[dvips]{graphics,color}
\usepackage{epsfig}\usepackage{float}
\RequirePackage{amssymb}
\usepackage{makeidx}
\usepackage[brazil]{babel}
\usepackage[latin1]{inputenc}
\usepackage{graphicx}
\usepackage{indentfirst}
\usepackage{color}
\newcommand{\be}{\begin{equation}}
\newcommand{\ee}{\end{equation}}

\newcommand{\ba}{\begin{eqnarray}}
\newcommand{\ea}{\end{eqnarray}}

\begin{document}

\title{ Brane Cosmic String Compactification in Brans-Dicke Theory }

\author{M. C. B. Abdalla$^{1}$}
\email{mabdalla@ift.unesp.br}
\author{M. E. X. Guimar\~aes$^{2}$}
\email{marg@unb.br}
\author{J. M. Hoff da Silva$^{1}$}
\email{hoff@ift.unesp.br}

\affiliation{1. Instituto de F\'{\i}sica Te\'orica, Universidade
Estadual Paulista, Rua Pamplona 145 01405-900 S\~ao Paulo, SP,
Brazil}

\affiliation{2. Departamento de Matem\'atica, Universidade de
Bras\'{\i}lia, Asa Norte 70910-900, Bras\'{\i}lia-DF, Brazil}


\begin{abstract}

We investigate an alternative compactification of extra dimensions
using local cosmic string in the Brans-Dicke gravity framework. In
the context of dynamical systems it is possible to show that there
exist a stable field configuration for the Einstein-Brans-Dicke
equations. We explore the analogies between this particular model
and the Randall-Sundrum scenario.

\end{abstract}
\maketitle

The search of an alternative compactification of extra dimensions
is a very interesting tool to analyze the hierarchy problem in
string theory. Among other physical implications they provide a
beautiful scenario of our universe insering it on a brane
\cite{Rubakov,Hamed}. In the Randall-Sundrum (RS) model there are
two 3-branes in a quite glance orbifold with $\mathbb{Z}_{2}$
symmetry in five dimensions \cite{RS}. A warp factor in the metric
shows how this type of model can help us with the hierarchy
problem, and beyond it, the model is a $5-D$ realization of
Horava-Witten solution \cite{HW}. Many physical effects related to
the extra dimensions can be calculated using the standard
brane-world gravity \cite{ROY}. However, from the point of view of
topological defects, these models use global domain walls to
generate the branes \cite{ovrut}. But cosmologists call our
attention to cosmological problems related to global defect like
domain walls \cite{vilenkin}. In this vein an interesting
alternative (global strings) was given in \cite{RUTH} that
provides in six dimensions all the necessary structure;
three-brane plus transverse space.

On the other hand, scalar-tensorial theories, in particular the
Brans-Dicke \cite{BD}, stand for many reasons as a good laboratory
to develop this kind of physical models. First, they show a robust
relation with gravity coming from fundamental string theory at low
energy. Second, they can be tested by experimental observations.
Besides, theoretically we hope that the additional scalar field
plays an important role in the distribution of masses of the
fundamental particles.


A particular and important characteristic of the models like RS
and the one found in \cite{cohen,RUTH} is that the topological
defect can compactify the spacetime around it by itself, leading
to a singularity. Topological domain walls and global strings
effectively do it, but if one introduces a time dependent factor
on the metric in the case of a global string the solution presents
a cosmological event horizon free of singularity \cite{R2}.
However this event horizon is not stable \cite{RJ}. In the
Einstein's theory only global defects have this property. However,
when we look at the equations of a local cosmic string in the
Brans-Dicke theory they seem to have that property, except for the
fact that there is no singularity at GUT scale \cite{CORDA}. In
other words, local cosmic string in the Brans-Dicke theory is
similar, in this context, to global string in Einstein's gravity.

Our effort here is to present a model of alternative
compactification using a local string in the Brans-Dicke theory.
In the course of development we shall use several approximations
and by using a dynamical system argument we arrive at a model
similar to the one in \cite{cohen,RUTH}. Nevertheless, we would
like  to stress that the main difference between the previous
works and ours is that we use local cosmic string in the framework
of a low energy string theory while they use a global string in
Einstein's theory. Besides, the space is topologically different
since we end up with a de Sitter behavior, and the additional
scalar field plays an important role in the warp factor.

We start with the equation \ba
R^{\mu}_{\nu}=2\partial^{\mu}\sigma\partial_{\nu}\sigma+\varepsilon
(T^{\mu}_{\nu}-\frac{\delta^{\mu}_{\nu}}{p+1}T)-\frac{2\delta^{\mu}_{\nu}\Lambda}{p+1},\label{1}
\ea which comes from the Brans-Dicke action in $p+3$ dimensions in
Einstein frame. In (\ref{1}), $\Lambda$ is the cosmological
constant, $\sigma$ is the scalar field, $T^{\mu}_{\nu}$ is the
stress tensor of a cosmic string, $T$ is the trace of the stress
tensor and $\varepsilon=-\frac{1}{M^{p+1}_{p+3}}$, where $M_{p+3}$
is the analogue of the Planck mass in $p+3$ dimensions. The
equations that make the relationship between the two frames (the
Jordan-Brans-Dicke and the Einstein frames) are
$w=\frac{1}{4\beta^{2}}-\frac{3}{2}$ (the Brans-Dicke parameter),
$\tilde{\phi}=\frac{1}{\varepsilon}e^{-2\beta\sigma}$ and
$\tilde{g}_{\mu\nu}=e^{2\beta\sigma}g_{\mu\nu}$. The physical
quantities (i.e., the quantities defined in the JBD frame) are the
tilde ones. Note that in the Einstein's frame the stress tensor is
not conserved.

In order to calculate the metric we expect that the spacetime has
the same symmetry of the source, and since our source is a cosmic
string, we write down the ansatz
\be{ds^{2}=e^{A}(-dt^{2}+dz_{i}^{2})+dr^{2}+e^{C}d\theta^{2}}\label{2},\ee
where $i=1...p$ and $A$ and $C$ are functions of $r$ only. So, all
the bulk structure has a cylindrical symmetry. Now we turn to tell
a little bit more about the source.

The standard model for a gauge cosmic string is the $U(1)$
invariant lagrangean \ba L&=&\left.
\frac{1}{2}(D_{\mu}\Phi)(D^{\mu}\Phi)^{*}+\lambda(\Phi^{*}\Phi-\eta^{2})\right.
\nonumber \\&+&\left.
\frac{F_{\mu\nu}F^{\mu\nu}}{16\pi}\right.\label{3},
 \ea
where $D_{\mu}=\partial_{\mu}+ieA_{\mu}$ and
$F_{\mu\nu}=\partial_{\mu}A_{\nu}-\partial_{\nu}A_{\mu}$. The
ansatz to generate a string-like solution is $\Phi=\eta
Xe^{i\theta}$ and $A_{\mu}=\frac{1}{e}(P-1)\partial_{\mu}\theta$.
Here $X$ and $P$ are also functions of $r$ only. Now, two
important remarks: first, when the masses of gauge and scalar
fields are equal, the fields have a behavior, far from the string
in Einstein theory, given by \cite{Linet}
 \ba
e^{C}\rightarrow r^{2}(1-4G\eta^{2})^{2}\nonumber, \\
P\rightarrow 2\sqrt{2}(1-4G\eta^{2})\gamma
r^{1/2}e^{-2\sqrt{2}r},\label{4}\\
X\rightarrow \gamma r^{-1/2}e^{-2\sqrt{2}r},\nonumber
 \ea
where $\gamma$ is a constant that can be determined by some
boundary condition. Besides, it is possible to show that in the
Brans-Dicke theory the stress tensor associated to (\ref{3}) in
Einstein frame is \cite{CORDA}
\begin{widetext}
\ba T^{t}_{t}=T^{z}_{z}=-\frac{\eta^{2}}{2}\Bigg(
X'^{2}e^{2\beta\sigma}+e^{-C}X^{2}P^{2}e^{2\beta\sigma}+2(X^{2}-1)^{2}e^{4\beta\sigma}+\frac{1}{8}e^{-C}P'^{2}\Bigg),\nonumber
\\
T^{r}_{r}=\frac{\eta^{2}}{2}\Bigg(X'^{2}e^{2\beta\sigma}-e^{-C}X^{2}P^{2}e^{2\beta\sigma}-2(X^{2}-1)^{2}e^{4\beta\sigma}+\frac{1}{8}e^{-C}P'^{2}
\Bigg), \label{5}
\\
T^{\theta}_{\theta}=\frac{\eta^{2}}{2}\Bigg(-X'^{2}e^{2\beta\sigma}+e^{-C}X^{2}P^{2}e^{2\beta\sigma}-2(X^{2}-1)^{2}e^{4\beta\sigma}+\frac{1}{8}e^{-C}P'^{2}
\Bigg),\nonumber
 \ea
\end{widetext} where prime means derivative with respect to r.

A remarkable characteristic of the solution found by Gundlach and
Ortiz \cite{CORDA} is that the string solution in Brans-Dicke
theory can be obtained using an amazing approximation far enough
from the core, so that (\ref{4}) is valid, in such way that
$e^{\beta\sigma}\simeq 1$ is still a good approximation. We shall
use it here. Note that in the components of stress-tensor it means
$\sigma \sim cte$, i.e. the contribution of a scalar Brans-Dicke
field, as a source, is too small compared with that one coming
from the cosmic string, even in the $r \gg$ region. Since we are
interested in a exotic compactification model, our system will be
analyzed in the $r \gg$ region. Taking everything into account, i.
e., substituting the expressions (\ref{4}) into (\ref{5}) and
using $e^{\beta\sigma}\simeq
 1$ we have, at $1/r$ order,
 \ba
T^{t}_{t}-\frac{T}{p+1}\simeq 16\Big( \frac{1-p}{1+p}
\Big)\eta^{2}\gamma^{2}\frac{e^{-4\sqrt{2}r}}{r},\nonumber\\
T^{r}_{r}-\frac{T}{p+1}=T^{\theta}_{\theta}-\frac{T}{p+1}\simeq
\frac{32\eta^{2}\gamma^{2}}{p+1}\frac{e^{-4\sqrt{2}r}}{r}.
\label{6}
 \ea
So, the Einstein-Brans-Dicke equation (\ref{1}), after some
calculation and calling $f(r)=\frac{32\varepsilon
\eta^{2}\gamma^{2}}{p+1}\frac{e^{-4\sqrt{2}r}}{r}$, gives
\begin{widetext}
\ba (p+1)\Big(A''+\frac{A'^{2}}{2}
\Big)+C''+\frac{C'^{2}}{2}\simeq
-4\sigma'^{2}-2f(r)+\frac{4\Lambda}{p+1},\label{7}\\
C''+\frac{C'^{2}}{2}+\frac{(p+1)}{2}A'C'\simeq
-2f(r)+\frac{4\Lambda}{p+1},\label{8}\\
A''+\frac{(p+1)}{2}A'^{2}+\frac{A'C'}{2}\simeq
f(r)(p-1)+\frac{4\Lambda}{p+1}.\label{9}
 \ea
\end{widetext}
It is easy to note that even in such approximation the equations
above are not simple. We shall perform, as in \cite{RUTH}, a
dynamical system formulation and then, an analysis in the phase
plane. We start with the variables \ba
x=pA'+C',\nonumber \\
y=C'. \label{10} \ea

The equations (\ref{7}) and (\ref{8}) together give ($f(r)\equiv
f$) \ba{-4\sigma'^2
=(p+1)\Bigg(A''+\frac{A'^{2}}{2}-\frac{A'C'}{2}\Bigg)}\label{e1},\ea
and substituting $A''$ from (\ref{9}) we have
\begin{eqnarray}
-4\sigma'^2 &=&\left.(p+1)\Bigg( (p-1)f+\frac{4\Lambda}{p+1}
\right. \nonumber \\&-&\left. \frac{p A'^{2}}{2}-A'C'
\Bigg)\right.\label{e3}.
\end{eqnarray}
Then, taking into account that
$\frac{x^{2}-y^{2}}{2p}=\frac{pA'^{2}}{2}+A'C'$, we arrive at \ba
4\sigma'^2&=&\left. -(p^{2}-1)f-4\Lambda \right. \nonumber
\\ &+&\left. \frac{(p+1)}{2p}(x^{2}-y^{2})\right.\label{11}.\ea
The dynamical system (\ref{10}) can be studied in the $(x,y)$
plane and the relation (\ref{11}) is a consistency equation that
plays an important role later on. From (\ref{8}) it is easy to see
that
\be{y'=-2f+\frac{4\Lambda}{p+1}-\frac{1}{2p}[xy(p+1)-y^{2}]}\label{12},\ee
and (\ref{7}), (\ref{9}) and (\ref{11}) led to \ba x'&=&\left.
(p+1)(p-2)f+4\Lambda\right. \nonumber
\\&-&\left. \frac{1}{2p}[(p+1)x^{2}-xy].\right.\label{13}
 \ea
The equations (\ref{13}) and (\ref{12}) form a non-autonomous
dynamical system that is characterized by two critical points \ba
(\bar x,\bar y)_{\pm}&=&\left. \pm \Delta
\Bigg(4\Lambda+f(p+1)(p-2)\right. \nonumber
\\ &,& \left. \frac{4\Lambda}{p+1}-2f\Bigg),\right. \label{14} \ea where
$\Delta=\Bigg(\frac{2(p+1)}{4\Lambda(p+2)+f(p^{2}-3)(p+1)}
\Bigg)^{1/2}$, and the $(\bar x,\bar y)_{+}$ and $(\bar x,\bar
y)_{-}$ are the attractor and repellor points, respectively.
Basically it means that there is, at least, one stable
configuration for the fields (attractor). Looking at (\ref{2}) one
notes that if $p=3$ we have a brane-world picture, with a
transverse space $(r,\theta)$. Beyond this, the attractor point in
the phase plane gives the integral equations to the fields in
$r\gg$, i. e.
\be{\sigma'^{2}=\frac{-4f(\Lambda+2f)}{5\Lambda+6f}}\label{15},\ee
\be{C'=\Big(\frac{2}{5\Lambda+6f}\Big)^{1/2}(\Lambda-2f)}\label{16},\ee
\be{A'=\Big(\frac{2}{5\Lambda+6f}\Big)^{1/2}(\Lambda+2f)\label{17}}.\ee

From equation (\ref{15}) we note that, since $f<0$ (because
$\varepsilon <0$), the cosmological constant is positive and must
obey the following inequality \be{\Lambda >2|f(r)|}\label{18}.\ee
It is a quite remarkable relation: First, it tells that we are, in
fact, dealing with a de Sitter space. Second, this  is a new
characteristic  of models of this kind. Like in reference
\cite{RUTH}, there are two branches and the system never goes to
the repellor. In a more explicit way, the equation (\ref{18})
gives \be{\Lambda
>16|\varepsilon|\eta^{2}\gamma^{2}\frac{e^{-4\sqrt{2}r}}{r}}\label{19},
\ee which leds to a very small cosmological constant.

The next step is to analyze how this model can help in solving the
hierarchy problem. Then, after coming back to the physical frame
we have
 \ba
ds^{2}=W(r)(\eta_{\mu\nu}dx^{\mu}dx^{\nu})+e^{2\beta\sigma}dr^{2}+H(r)d\theta^{2}\nonumber,\ea
where $W(r)=e^{2\beta\sigma+A}$, $H(r)=e^{2\beta\sigma+C}$ and
 $A$, $C$ and $\sigma$ are the fields which asymptotic behavior is
found in (\ref{15})-(\ref{17}). Explicitly, the function $W(r)$
intervenes directly in the masses generated by the Higgs
mechanism, just like in \cite{RS}. Nevertheless, we should
emphasize here that there is a new possible adjustment provided by
the Brans-Dicke scalar field. However, the way it appears can not
tell, in a rigorous way, its influence in the Higgs mechanism once
the equations are only valid far from the brane. By all means, it
is clear that inclusion of the Brans-Dicke field opens a new
possibility.

It is easy to note from the equations (\ref{15})-(\ref{17}) that
if one implements the limit $f(r)\rightarrow 0$, the scalar field
becomes constant, and the fields $A$ and $C$ are equal, apart of a
constant. When this occurs the functions $W(r)$ and $H(r)$ become
equal and we arrive into a new brane-world picture, given by \ba
ds^{2}=e^{k\Lambda^{1/2}r}(\eta_{\mu\nu}dx^{\mu}dx^{\nu}+d\theta^{2})+dr^{2}\label{20},\ea
where $k$, in general, depending on the dimension and the
constants had been absorbed in new variables $(x^{\mu},\theta,r)$.
Again, the equation (\ref{20}) is valid only in $r\gg$ region (far
from the string)  and we stress the minuteness of the cosmological
constant. Of course, the shape found in (\ref{20}) can also be
obtained solving the Einstein's equations with cylindrical
symmetry without any source and free of approximations, however
there is no physical reason to do it.

We conclude this work stressing that we believe that this  is a
promising field of research. I.e., to realize brane-gravity on
this model, analyzing how the extra-dimensions influence in the
calculation of measurable physical amounts \cite{LHC}. Especial
techniques used in other models cannot be used here, like among
others, Israel junction conditions \cite{Israel} at the brane that
makes explicitly use of the $\mathbb{Z}_{2}$ symmetry.

The combination of a local cosmic string and the low energy string
theory such as Brans-Dicke seems to be a good alternative model
for compactifying extra dimensions from both physical and
topological points of view.

The authors would like to thank Profs. A. A. Bytsenko and J. A.
Helayel-Neto for useful comments on this manuscript. J. M. Hoff da
Silva would like to acknowledge CAPES-Brazil for financial
support. M. E. X. Guimar\~aes and M. C. B. Abdalla acknowledge
CNPq for a support.

\end{document}